\documentclass[12pt]{article}
\usepackage{authblk}
\usepackage[english]{babel}
\usepackage{braket}

\usepackage[letterpaper,top=2cm,bottom=2cm,left=2.5cm,right=2.5cm,marginparwidth=1.75cm]{geometry}

\usepackage{amsmath}
\usepackage{bbold}
\usepackage{graphicx}
\usepackage[colorlinks=true, allcolors=blue]{hyperref}
\usepackage{setspace}
\title{Born-Oppenheimer and the Geometry of Ray Space:\\ an application to cold atoms}
\author{Joseph Samuel}
\affil{International Centre for Theoretical Sciences, Bangalore, India 560089\newline
and\\Raman Research Institute, Bangalore, India 560080}
\begin{document}
\maketitle
\doublespacing
\begin{abstract}
It is known that, within the Born-Oppenheimer
approximation, the slow modes of the nuclear motion are  
altered by three effects that emerge from integrating out the fast modes
of the electronic motion. The first is
an effective scalar potential $V_{\mathrm dyn}$ coming from the eigenvalue of the electronic state, the second is 
an effective magnetic field coming from the Berry phase vector potential $A$. The third term   
is an additional potential $V_{\mathrm geom}$ originating in the geometry of ray space and the Fubini-Study metric. 
In this article, we illustrate these effects and their geometric origin 
in the context of a simple toy model of a slow neutron interacting with a strong, 
spatially varying magnetic field. 
Regarding the neutron spin  as a fast degree  of freedom, we work out the slow
dynamics of the motion of the neutron. Our treatment is geometrical and brings out the effects
originating in the K\"ahler geometry of ray space and the Fubini-Study metric.
We then give examples of magnetic field configurations which isolate these three separate effects. 
Finally we apply these ideas to the trapping of cold atoms. Our main result is that the
geometric electric potential $V_{\mathrm geom}$ dominates for smaller traps and 
can be used to confine cold atoms in static traps.
This observation could result in better and smaller atomic clocks. 
This paper is dedicated to Michael Berry on
his 80th Birthday.
\end{abstract}
\date{email: sam@icts.res.in, sam@rri.res.in}\\
\vspace{3cm}
\newline
\vspace{10.0in}

\newpage

\section{Introduction}
Berry's work\cite{berryprocroysoc} has illuminated the geometry underlying the ray space of quantum mechanics. 
The seminal ideas in \cite{berryprocroysoc} led to numerous experiments (for example, \cite{chiaowu,bhandarisamuel,poles}), applications and generalisations\cite{wilczekshapere,aharonovanandan}. 
It was also realised that the idea had been anticipated several times\cite{panch,ramaseshannityanand} in different contexts. 
Michael Berry traces the seed of the idea as far back as William Rowan Hamilton (1832) in his theoretical studies on conical double refraction
and their experimental confirmation by Herbert Lloyd. 
One of the anticipations was due to Longuet-Higgins\cite{longuethiggins} in the context of
the Born-Oppenheimer approximation which is widely used in the quantum theory of lattice dynamics and molecular physics. 
The Born-Oppenheimer approximation
separates the fast electronic motion from the slow nuclear motion. Initially the motion of the nucleii is neglected and the frozen nuclear
coordinates are treated as parameters in the electronic problem. Invoking the adiabatic theorem to allow for slow motion of the nuclei,
one then uses a  reduced Schr\"odinger equation to solve for the nuclear motion.  The purpose of this article is to elucidate the 
derivation of the slow dynamics in a simple toy model. While this topic has been studied before in Refs.\cite{ZYGELMAN1987476,MVBerry1990,berryfive},
our treatment may perhaps appeal to a geometrically minded reader. We go on to apply these geometric ideas to cold atom 
physics and make suggestions which could potentially result in better atomic clocks.

 Let us motivate our discussion with a simple classical example . Consider a well oiled bicycle wheel\footnote{The precession
and nutation of a rapidly spinning top can also be viewed as slow modes subject to a monopole magnetic field in exactly the same way.} mounted on an axle rod so that it is free to rotate
 about an axis pointing along the rod. If the direction of the rod is held fixed, and the wheel spun at angular velocity $\omega$ 
 the wheel rotates through an angle $\omega t$ in time $t$ (neglecting friction).  If the direction of the 
 rod is slowly changed  to traverse a closed circuit $C$ in the space of directions, we may naively expect that the total angle of rotation is
 still $\omega t$. However, this naive expectation is wrong. The total angle of rotation includes an additional piece given by the solid angle traversed
 by the circuit $C$. The explanation is that the wheel rotates about different axes during the circuit and such rotations do not commute. 
 This effect is the Hannay angle\cite{hannay}, a classical analogue of the Berry phase. 
 To perform this experiment, it helps to mark  the rim of the bicycle wheel and use a stroboscope or camera
 to track the motion of the wheel. 

So far we have regarded the direction of the rod as a parameter within the control of the experimenter. 
However, parameters are also dynamical variables, subject to the laws of motion. 
To realise this, let us suspend the wheel and axle 
from the ceiling and repeat the experiment by spinning the wheel and setting the axle pendulum in motion. 
We find that the trajectory of the axle resembles that of  a charged particle in a magnetic field. 
 With more work one can arrange for the axle to trace its path in a sand pit and see 
precessing ellipses. Reversing the sense of the spin reverses the sense of the precession. 
What we are seeing here, in the spirit of the Born-Oppenheimer approximation, is the back reaction of the fast mode 
(the spin of the bicycle wheel) on the slow modes (the oscillation of the pendulum) 
of the system. A less elaborate
experiment, less convincing, but more in tune with the sloppy habits of theoretical physicists, is to twist a wet 
tea bag so its string is wound up and let it untwist while
oscillating, suspended from its string. 
Michael Berry has been subjected to both these demonstrations on one of his visits to the Raman Research Institute.

In this article, we consider a simple quantum mechanical toy model\footnote{ Another toy model has been discussed in  \cite{gangopadhyayduttaroy}} to study the Born-Oppenheimer
approximation. This quantum model is described in section II. In section III, we derive the dynamics of the slow modes
taking into account the back reaction of the fast mode. In section IV
we describe some geometrical aspects of ray space to bring out the geometric nature
of the extra scalar potential. Section V describes magnetic field configurations which  bring out the 
effect of different terms in the slow Hamiltonian. Section VI shows how the geometrical ideas of the toy model carry over to the more
realistic model of molecular dynamics. In section VII, we suggest that geometric electromagnetic fields may find application in
developing better atomic clocks. Section VIII is a concluding discussion.

{\it Notation:} We use Dirac's braket notation for the fast mode but not for the slow modes. Projectors onto one dimensional subspaces (proportional to the vector $\ket{e}$)
of the fast Hilbert space
are denoted $P^e$ and satisfy $P^e P^e=P^e$. The orthogonal projection is $Q^e=\mathbb{I}-P^e$. We will use a global space independent frame $\{\ket{+},\ket{-}\}$
as well as a local frame $\{\ket{e_+(\vec{R})},\ket{e_-(\vec{R})}\}$, which is space dependent. Spatial derivatives, $\frac{\partial}{\partial R^i}$ will be denoted $\nabla_i$.  
The projectors $P^+$ and $Q^+=P^-=\mathbb{I}-P^+$, commute with $\nabla_i$, since the basis 
$\{\ket{+},\ket{-}\}$ is independent of space. Let us also clarify here that there are two kinds of electromagnetic effects in play here.
Those electromagnetic fields that are produced by charges and currents are called laboratory electromagnetic fields.
Those electromagnetic fields that emerge from the geometry of ray space are referred to as geometric electromagnetic fields
(also referred to in \cite{lin} as ``synthetic gauge fields'').  

\section{A Toy Model}
We consider a slow neutron in a strong  external magnetic field ${\vec{B}}({{\vec{R}}})$. 
Let us regard the spin as the fast variable and
the spatial position ${\vec{R}}$ of the neutron as the slow variable.
The total Hamiltonian of our system is the sum  
\begin{equation}
 H_T=H_f+H_s \hspace{.5cm}.
\label{hamiltonian}
\end{equation}
The fast Hamiltonian $H_f$ is that of a spin half in a magnetic field
\begin{equation}
H_f=-\mu {\vec B}({\vec{R}}).{\vec{\sigma}}\hspace{.5cm},
    \label{fasthamiltonian}
\end{equation}
where ${\vec B}({\vec{R}})$ is a spatially 
varying magnetic field and $\mu$ is the magnetic dipole moment of the neutron.
We will assume that $|\vec{B}(\vec{R})|$ does not vanish in the region of interest. In the spirit of the Born-Oppenheimer approximation, we will assume that
the magnetic field $\vec{B}(\vec{R})$ varies slowly with ${\vec{R}}$. (More precisely, we would like 
$\frac{d\vec{R}}{dt}.\vec{\nabla}\big(\hat{B}(\vec{R})  \big)<<\frac{\mu |\vec{B}(\vec{R})|}{\hbar}$, where $\hat{B}$ is a unit vector 
and $\frac{d\vec{R}}{dt}$ a typical velocity of the slow neutron
.)

The slow Hamiltonian $H_s$
is given by the kinetic energy of the neutron
\begin{equation}
H_s=- \frac{\hbar^2}{2 m}{\vec \nabla}.{\vec \nabla}\hspace{.5cm}.
\label{slowhamiltonian}
\end{equation}
We now regard ${\vec{R}}$ as a parameter in the fast Hamiltonian
$H_f(\vec{R})$ and diagonalise the fast Hamiltonian. (In molecular physics, this corresponds to holding the nuclear configuration fixed and solving for the
electronic eigenstates.)
Let the total wave function of the system be
\begin{equation}
\Psi({\vec{R}})=\begin{bmatrix}
           \psi_+(\vec{R}) \\
           \psi_-(\vec{R})
         \end{bmatrix}=\psi_+({\vec{R}}) \ket{+}+\psi_2({\vec{R}}) \ket{-},
    \label{wavefunction}
\end{equation}
where 
\begin{eqnarray}
\ket{+}&= \begin{bmatrix}
           1 \\
           0
         \end{bmatrix} ,\,\,\,
          \ket{-}&= \begin{bmatrix}
           0 \\
           1
         \end{bmatrix}
         \end{eqnarray}
We will refer to the orthonormal basis $\{\ket{+},\ket{-}\}$
as the global frame. We will also need a local frame adapted to the magnetic field. To do this, 
let us solve the eigenvalue equation
 \begin{equation}
 H_f(\vec{R}) \ket{e({\vec{R}})}=\lambda\ket{e(\vec{R})}
     \label{eigenfast}
 \end{equation}
 for the fast spin degree of freedom. 
\begin{sloppypar}
 This equation has 
 two independent solutions $\{\ket{e_+(\vec{R})},\ket{e_-(\vec{R})}\}$ 
 (satisfying $\braket{e_+(\vec{R})|e_-(\vec{R})}=0$)
 corresponding to the two distinct eigenvalues $\{\lambda_+=-\mu |B|\},\lambda_-=\mu |B|\}$. For normalised eigenstates $\ket{e_{\pm}(\vec{R})}$,
 equations (\ref{eigenfast}) determine $\ket{e_a(\vec{R})}, a=\{+,-\}$ up to a phase 
factor $w_a(\vec{R}),|w_a(\vec{R})|=1$ a complex number of modulus unity, which depends on $a$ as well as $\vec{R}$. Let us make
an arbitrary smooth choice of these phases\footnote{This is always possible locally.}.  
We now have $\{\ket{e_+(\vec{R})},\ket{e_-(\vec{R})}\}$ , which defines a local orthonormal basis which is adapted to the magnetic field and provides
us with a moving frame. We refer to this as the local basis. 
\end{sloppypar}

The two orthonormal bases $\ket{a}$ and $\ket{e_a(\vec{R})}$ are related by a Unitary transformation $U(\vec{R})$
\begin{equation}
    \label{unitary}
    \ket{e_a(\vec{R})}=U(\vec{R}) \ket{a}
\end{equation}
 We can transform our original eigenvalue problem (\ref{eigenfast}) by defining
 $\Psi(\vec{R})=U(\vec{R}) \Phi(\vec{R})$.  The new eigenvalue equation is 
 \begin{equation}
 {\cal H}_T \Phi= E \Phi,
 \label{neweigenvalue}
 \end{equation}
 where ${\cal H}_T=U^{-1}  H_T U$ is given by
 \begin{equation}
 {\cal H}_T=\lambda_+(\vec{R}) \sigma_3-\frac{\hbar^2}{2m} U^{\dagger}(\vec{R})\nabla_i \nabla_i U(\vec{R})
     \label{hprimeexpression}
 \end{equation}
 Let us now work out the dynamics governing the slow modes of the system.
 
\section{The Reduced Hamiltonian}
Let us assume that the spin is in the ground state $\ket{+}$ of the fast Hamiltonian. The adiabatic theorem assures us that (with an exponentially small chance of transition) the spin remains in the ground state. We can now pass to the reduced space by projecting the wave
function and the Hamiltonian into the $\ket{+}$ subspace. $\phi(\vec{R})=\braket{+|\Phi(\vec{R})}$. 
The spin state can be described by the density matrix $\rho_f=(1+\sigma_3)/2=\ket{+}\bra{+}=P^+$, where $P^+$ is the one dimensional projector on to the ground state. The reduced Hamiltonian for the slow degrees of freedom is ${\cal H}_{red}=Tr[{\cal H}_T \rho_f]=\bra{+}{\cal H}_T\ket{+}$. 

Writing out the reduced Hamiltonian, we have for any wave function $\xi(\vec{R})$
\begin{equation}
{\cal H}_{red}\xi(\vec{R})=\lambda_+(\vec{R}) \braket{+|\sigma_3|+}\xi(\vec{R})-\frac{\hbar^2}{2m} \bra{+}U^{\dagger}(\vec{R})\nabla_i \nabla_i U(\vec{R})\ket{+}\xi(\vec{R}),
    \label{hreduced}
\end{equation}
where it is understood that the derivative operators act on everything to their right unless explicitly bracketed. 
\subsection{Dynamical Scalar Potential}
The first term in ${\cal H}_{red}$, gives us the fast eigenvalue which gives an effective space dependent scalar potential $V_{dyn}(\vec{R})=-\mu |\vec{B}(\vec{R})|$ for the slow modes. 
We call this the dynamical scalar potential as it is related to the eigenvalue of the fast Hamiltonian. 
This term is well known in molecular physics: the electronic eigenvalue serves as a scalar potential in which the slow
nucleii move. Adiabatically moving a neutron to a region of lower magnetic field requires the mover to do some work. 
\subsection{Geometric Vector Potential}
The second term in ${\cal H}_{red}\xi(\vec{R})$ can be rewritten as
\begin{equation}
-\frac{\hbar^2}{2m} \braket{+|U^{\dagger}(\vec{R})\nabla_i U(\vec{R}) U^{\dagger}(\vec{R})\nabla_i U(\vec{R})|+}\xi(\vec{R})
    \label{kinetic}
\end{equation}
by inserting the identity $U U^{\dagger}=1$ between the derivatives. It 
is elementary to verify that for any function $\xi(\vec{R})$
\begin{equation}
U^{\dagger}(\vec{R}) \nabla_i U(\vec{R})\,\,\xi(\vec{R})=\nabla_i\xi(\vec{R})+\omega_i (\vec{R})\xi(\vec{R})    
\label{ym}
\end{equation}
where $\omega_i(\vec{R})=U^{\dagger}(\vec{R}) \nabla_i U(\vec{R})$ is an effective $U(2)$ gauge field (with vanishing $U(2)$ field strength). 
Writing ${\cal D}_i=\nabla_i+\omega_i$ as the covariant derivative in the $U(2)$ gauge field, we have 
\begin{equation}
    \label{identity}
Tr[P^+{\cal D}_i {\cal D}_i ]=\braket{+|{\cal D}_i {\cal D}_i|+}=\braket{+|{\cal D}_i|+}\braket{+|{\cal D}_i|+}+
\braket{+|{\cal D}_iP^- {\cal D}_i|+},
\end{equation}
where $P^-=(\mathbb{1}- P^+)$.
For any function $\xi(\vec{R})$, 
$\braket{+|{\cal D}_i|+}\xi(\vec{R})=$
\begin{eqnarray}
    \nabla_i \xi(\vec{R})+\braket{+|\omega_i|+} \xi(\vec{R})&=
    \nabla_i \xi(\vec{R}) +[\braket{+|U^\dagger(\vec{R}) \nabla_i U(\vec{R})|+}]\xi(\vec{R})=\\ 
    \nabla_i \xi(\vec{R})+[\braket{e_+(\vec{R})|\nabla_i|e_+(\vec{R})}]\xi(\vec{R})&=\nabla_i \xi(\vec{R})+A_i\xi(\vec{R})
    \label{berrypot}
\end{eqnarray} 
where $A_i=\braket{e_+(\vec{R})|\nabla_i|e_+(\vec{R})}$ is the Berry potential, which
we also refer to as $A_{geom}$ the geometric vector potential. As a result the first term in (\ref{identity})
gives us 
\begin{equation}
(\nabla_i+A_i) (\nabla_i+A_i)\xi(\vec{R})
    \label{first}
\end{equation}
This term shows that the slow modes are subject to an effective 
magnetic field represented by Berry potential $A_i$. This is a well known 
result, which has been discussed in the molecular physics literature\cite{meadtruhlar}.
This is the quantum analogue of the classical effect described in the introduction of this paper. 
\subsection{Geometric Scalar Potential}
Like the Berry potential, the second term in (\ref{identity}) also 
has a simple geometric interpretation.
Expanding the expression for the covariant derivative, 
we have
\begin{equation}
\braket{+|(\nabla_i +\omega_i) P^- (\nabla_i+\omega_i)|+}
    \label{expansion}
\end{equation}
Since $\nabla_i$ commutes with $P^-$, all the derivative terms drop out in (\ref{expansion}) leaving us with 
$\braket{+|\omega_i P^-\omega_i|+}=-\braket{+|\omega_i^\dagger P^- \omega_i|+}$ which equals
\begin{eqnarray}
    \label{final}
    -\braket{+|\big(\nabla_i U^\dagger(\vec{R})\big) U(\vec{R}) P^- U^\dagger(\vec{R}) \big(\nabla_i U(\vec{R})\big)|+}
    =-\big(\nabla_i\braket{e_+(\vec{R})|\big)U(\vec{R}) P^- U^\dagger(\vec{R})\big(\nabla_i|e_+(\vec{R})}\big)\\
    =-\big(\nabla_i\bra{e_+(\vec{R})}\big)(\mathbb{I}-
    \ket{e_+(\vec{R})}\bra{e_+(\vec{R}))\big(\nabla_i|e_+(\vec{R})}\big)=
    -\big(\nabla_i\braket{e_+(\vec{R})|\big)
    Q^{e_+}\big(\nabla_i|e_+
    (\vec{R})}\big)
\end{eqnarray}
where $Q^{e_+}=P^{e-}=\mathbb{I}-P^{e+}$ is the projector orthogonal to  the state $\ket{e_+(\vec{R})}$.
Supplying the necessary constants, we find the final expression for the geometric scalar potential
\begin{equation}
    \label{geometricscalarpotential}
    V_{geom} (\vec{R})=\frac{\hbar^2}{2m}{\huge\{  } \big(\nabla_i\braket{e_+(\vec{R})|\big)\big(\nabla_i|e_+(\vec{R})}\big)-\big(\nabla_i\braket{e_+(\vec{R})|\big)
    \ket{e_+(\vec{R}}\bra{e_+(\vec{R}}\big(\nabla_i|e_+(\vec{R})}\big){\huge\} },
    \end{equation}
This potential is non-negative and vanishes only when the {\it direction} of $\vec{B}$ is constant\cite{MVBerry1990}. 
In the next section we will bring out the geometry behind these terms in the slow Hamiltonian.
\section{Geometry of Ray Space}
The ray space of quantum mechanics ${\cal R}$ is defined as the set of normalised states modulo multiplication by a unit complex number, a phase.
For finite dimensional Hilbert spaces, this is just complex projective space $\mathbb{CP}^n$. In the toy model discussed here $n=1$ and we have $\mathbb{CP}^1=S^2$. 
${\cal R}$ has three intertwining structures on it: a Riemannian structure, a symplectic structure and a complex structure. These three structures
combine to form a K\"ahler structure. 

The Riemannian structure is physically accessible as the overlap between quantum states:
$|\braket{\phi|\psi}|^2$ represents a transition probability. Geometrically, the same quantity gives us 
the distance $d$ between rays $|\braket{\phi|\psi}|=\cos{d/2}$. Passing to infinitesimally separated rays $\ket{\phi}=\ket{\psi}+\delta\ket{\psi}$, we arrive 
at the Fubini-Study metric \cite{eguchi,berryfive} for normalised states.
\begin{equation}
ds^2=\braket{\delta \psi|\delta \psi}-\braket{\delta \psi|\psi}\braket{\psi|\delta \psi}
    \label{fubinistudy1}
\end{equation}
where we choose representative states from each ray. Given a normalised $\ket{\psi}$ and 
two vectors $\delta_1\ket{\psi}$ and $\delta_2\ket{\psi}$,
the Fubini-Study metric is
\begin{equation}
    \label{fs}
    \gamma(\delta_1\ket{\psi},\delta_2\ket{\psi})=\braket{\delta_1 \psi|\delta_2 \psi}-\braket{\delta_1 \psi|\psi}\braket{\psi|\delta_2 \psi}
\end{equation}
The geometric scalar potential 
(\ref{geometricscalarpotential}) is exactly of this form with $\delta\ket{\psi}$ replaced by $\nabla_i\ket{e_+(\vec{R})}$.

Although we have used representative kets instead of rays, this is actually a metric on the ray space.
This metric can also be expressed directly in terms of density matrices. If $\rho$ is a density matrix and $\dot{\rho}$ and $\rho'$ tangent
vectors at $\rho$, the Fubini-Study metric is given by 
\begin{equation}
    \label{densitymetric}
    \gamma(\dot{\rho},\rho')=\frac{1}{2} \mathrm{Tr}[\dot{\rho}\rho']
\end{equation}

The symplectic structure is physically manifested in the geometric phase which is detectable in interference phenomena. The symplectic form on ray space is the curvature of the Berry
connection and acts like a magnetic field. Given any closed curve $C$ in ray space, the Berry phase is $\exp{i \phi}$, where $\phi$ is the symplectic area of any disc that $C$ bounds.
The Berry phase for  open curves can be similarly defined
(provided the ends of the curve are not on orthogonal rays) by closing the ends by a
geodesic curve. The geodesic in ray space is determined by the Fubini-Study metric. The work
of Pancharatnam\cite{panch} in optics drew attention to the importance of geodesics. These ideas were carried over to the Hilbert space of quantum mechanics in \cite{samuelbhandari}.

The complex structure is intimately
tied up with the symmetry of time reversal.  Time reversal is an anti-unitary operator in quantum mechanics. Wigner's theorem tells us that all 
ray space isometries are either unitary or anti-unitary. To discriminate between these two possibilities we use the Bargmann invariant \cite{bargmann,mukundasimon}
which is an example of the Pancharatnam phase\cite{panch}. For Hamiltonians which are 
time reversal invariant, the geometric phase can only take values $\pm1$ and 
becomes a topological phase. This was the context for Longuet-Higgins discovery
of conical intersections in molecular physics.

Let us now understand the geometric scalar potential in terms of the Fubini-Study metric. 
At every point $\vec{R}$, Eq. (\ref{eigenfast}) determines a ray in 
the fast Hilbert space. In the case of the two state sytem we consider here
the ray space is $S^2$, the Bloch sphere.  
\begin{equation}
f:\mathbb{R}^3\mapsto S^2
    \label{map}
\end{equation}
The map $f$ can be used to pull back the Fubini-Study metric tensor to real space. This gives us a second rank tensor in real space.
Contracting this tensor with the spatial metric gives us the geometric potential. Using the unit vector $\hat{n}$,
where $\rho(\vec{R})=\frac{1+\hat{n}(\vec{R})}{2}$, to describe a point on the Bloch sphere,
$f$ gives us $\hat{n}({\vec{R}})$ and we have the pull back explicitly written as
\begin{equation}
\gamma_{ij}=\frac{1}{2} \mathrm{Tr}[\big(\nabla_i\rho(\vec{R})\big)\big(\nabla_j \rho(\vec{R})\big)]=\frac{1}{4} (\nabla_i \hat{n}(\vec{R}))\cdot(\nabla_j\hat{n}(\vec{R})),
    \label{pullback}
\end{equation}
The potential $V_{geom}$ is then given by  $g^{ij}(\vec{R}) \gamma_{ij}(\vec{R})$ the contraction of $\gamma_{ij}$ with the metric tensor of space. Although $g_{ij}=\delta_{ij}$ is flat in this simple toy model, we include $\vec{R}$ dependence to leave open the possibility of using generalised coordinates on the slow configuration space.  The Fubini-Study metric is a Riemannian metric on the ray space and so is positive definite. 
However, the pull back $\gamma_{ij}$ of this tensor is not a Riemannian metric on space since it may be degenerate. It is however, non-negative (positive semi definite). Similarly, the pullback of the symplectic form on the ray space, gives us a two-form on space, but this could be 
degenerate (or indeed vanish entirely).

\section{Isolating the separate effects} 
Putting together the work of the last sections, we find the Hamiltonian for the slow modes:
\begin{equation}
    \label{slowmodehamiltonian}
    {\cal H}_{red}=V_{dyn}(\vec{R})-\frac{\hbar^2}{2m} (\nabla_i-A_i(\vec{R})) (\nabla_i-A_i(\vec{R}))+ V_{geom}(\vec{R}),
\end{equation}
where $V_{dyn}(\vec{R})=\lambda_+(\vec{R})$, $A_i$ is the Berry potential and $V_{geom}(\vec{R})$ is the geometric potential. 
We see that the slow modes are subject to three effects by the fast modes. 
\begin{enumerate}
    \item The first is a dynamic scalar potential $V_{dyn}(\vec{R})$ arising from the eigenvalue of the fast mode.
\item The second is modification of the kinetic
energy term by the magnetic field $F=dA$ of the Berry potential. Let us be clear that this is not the same as the external  $\vec{B}(\vec{R})$ we introduced; remember the neutron is neutral and insensitive to the external magnetic field $\vec{B}(\vec{R})$.  Only the spin couples to $\vec{B}(\vec{R})$. What we are seeing here is {\it geometric
magnetism}, which is distinct from the magnetic effects created by currents.
\item The third is the scalar geometric potential $V_{geom}$ that is of interest in this paper. One could think
of this as a ``{\it geometric electricity}'' in analogy with the usual scalar potential in electrostatics. 
\end{enumerate}
Let us look at some examples to isolate these separate effects.
By choosing a magnetic field $\vec{B}(\vec{R})= b(\vec{R}) \hat{i}$ which varies only in  magnitude, but not in direction, we eliminate effects 2 and 3
and isolate the dynamic scalar potential. To isolate the effect of the geometric  vector potential,
we consider the magnetic field $\vec{B}(\vec{R})=B_0(\hat{i} \cos{y}+\hat{j}+\hat{k} \sin{y})$ which has constant dynamic and
geometric potentials. Here effects 1 and 3 are eliminated and only effect 2 remains.

To isolate the effect of the geometric scalar potential, we present an example in which the 
first two effects are absent and the third is present. This will convince the reader that the third effect is independent of the first two. It may
also help an experimental detection of the third effect if the first two effects are absent. We will now describe a magnetic field configuration
$\vec{B}(\vec{R})$, (with $\vec{\nabla}.\vec{B}=0$) with the property that 
\begin{itemize}
\item $|\vec{B}(\vec{R})|$ is constant and so $V_{dyn}(\vec{R})$ is independent of $\vec{R}$.
\item the ``magnetic field'' $F=dA$ due to the Berry potential vanishes.
\end{itemize}

With $\vec{R}=(x,y,z)$,
consider $\vec{B}(\vec{R})=B_0(\cos{a(y)}\hat{i}+\sin{a(y)}\hat{k})$ where $a(y)$ is any function depending only on $y$. 
This makes the fast Hamiltonian $H_f=-\mu \vec{B}(\vec{R}).\vec{\sigma}$ real 
and proportional to $\cos{a(y)}\sigma_3+\sin{a(y)} \sigma_1$, which is 
\begin{equation}
\begin{bmatrix}
\cos{a(y)}&\sin{a(y)}\\
\sin{a(y})& -\cos{a(y)}\\
\end{bmatrix}
\end{equation}
The ground state of this Hamiltonian $H_f$ is 
\begin{equation}
    \label{gs}
    \ket{e_+(y)}=
    \begin{bmatrix}
    \sin{a(y)/2}\\
    -\cos{a(y)/2}\\
    \end{bmatrix}
\end{equation}
It is easily checked that 1) $\lambda_+(\vec{R})=-\mu B_0$ is actually independent of $\vec{R}$, so this term has no  effect on the slow modes.
2) The Berry potential vanishes identically $A_i=0$ and so there are no ``geometric  magnetic''  effects and 3) the geometrical scalar potential $V_{geom}$ does {\it not} 
vanish and is given by $\frac{\hbar^2}{8m} (\frac{da(y)}{dy})^2$. Note that $V_{geom}$ is non-negative. 
\section{More Realistic Models}
Let us carry these ideas over to the more realistic situation of  molecular dynamics. The total Hamiltonian
is 
\begin{equation}
    \label{molecular}
    H_T=T_N+V_{NN}+ T_{e}+V_{ee}+V_{Ne}
\end{equation}
where the terms are, in order, $T_N$ the nuclear kinetic energy, $V_{NN}$ the nuclear-nuclear interaction, 
$T_e$ the electronic kinetic energy, $V_{ee}$ the electron-electron interaction and $V_{Ne}$ the nuclear-electronic interaction.
The fast Hamiltonian consists of all those terms that depend on electronic coordinates,
\begin{equation}
    \label{hfast}
    H_f= T_{e}+V_{ee}+V_{Ne}(R),
\end{equation}
where it is understood that the dependence of $V_{Ne}$ on the nuclear coordinates $R$ is parametric. We use the symbol
$R$ to represent a set of generalised coordinates in the nuclear configuration space. The slow Hamiltonian consists of the remaining terms
\begin{equation}
    \label{hslow}
  H_s=  T_N+V_{NN}
  \end{equation}
Following the Born-Oppenheimer approach, we need to solve eq.(\ref{eigenfast}) for the eigenstates $\ket{e_n(R)}$ and eigenvalues $\lambda_n(R)$. 
While this problem is considerably more difficult than our simple toy model, it is within the reach of physical chemists aided by modern computers. 
We will assume that there are no degeneracies $\lambda_n(R)\ne \lambda_m(R)$ for $m,n$ distinct. 
The general ideas we used for the toy model carry over 
to these more realistic and difficult models. 
After performing a unitary transformation to the basis $\{\ket{e_n(R)}\}$, the slow mode Hamiltonian, including the back reaction of the fast modes on the slow ones is given by
\begin{equation}
    \label{hback}
    {\cal H}_{red}=\Tilde{T}_N+V_{NN}+V_{dyn}+V_{geom},
\end{equation}
where $\Tilde{T}_N$ is the nuclear kinetic energy modified by the Berry potential $A$, and $V_{dyn}$ and $V_{geom}$
are the two scalar potentials of dynamical and geometric origin. Assuming that the fast system is in the state $\ket{e_n(R)}$ with eigenvalue $\lambda_n(R)$,
here are explicit 
expressions for these 
\begin{equation}
    \label{vdyn}
    V_{dyn}(R)=\lambda_n(R).
\end{equation}
The effect of the geometric vector potential and the geometric scalar potential can be expressed in terms of the second rank
tensor ${\cal V}_{ij}=\overline{{\cal V}}_{ji}$ on the nuclear configuration space
\begin{equation}
\label{vfield}
    {\cal V}_{ij}(R)=\sum_{m\neq n} \frac{\braket{e_n(R)|\partial_{i} H_f(R)|e_m(R)}\braket{e_m(R)|\partial_{j} H_f(R)|e_n(R)}}{(\lambda_n(R)-\lambda_m(R))^2}
    \end{equation}
    The effective magnetic field strength due to the Berry potential is given by
    the imaginary part of the tensor ${\cal V}$:
    \begin{equation}
        \label{magfield}
         F_{ij}(R)=\frac{{\cal V}_{ij}-{\cal V}_{ji}}{2i}.
        \end{equation}
This agrees with eq.(10) of Ref. \cite{berryprocroysoc}. 
    The pull-back of Fubini-Study metric on the fast ray space to the nuclear configuration space is given by the real part:
    \begin{equation}
        \label{fubinistudypullback}
    \gamma_{ij}(R)=\frac{{\cal V}_{ij}+{\cal V}_{ji}}{2}
    \end{equation}
    The geometric scalar potential is
    given by
    \begin{equation}
    \label{geompot}
    V_{geom}(R)=g^{ij}(R) \gamma_{ij}(R),
    \end{equation}
where the metric $g$ on the nuclear coordinate space has been contracted with  $\gamma$. Note also that the metric on the nuclear configuration space
is determined by the kinetic energy term, which weights the nucleii according to their masses. 
It may be advantageous to leave open the choice of coordinates on this space. For instance nuclear separations
may be better coordinates than nuclear positions. 

\section{Better Clocks from Geometric Fields}
Many laboratories over the globe cool and trap atoms and ions. 
This field has overlap with condensed matter physics, quantum simulation 
and quantum computation. An important motivation for cold atom studies is to achieve more accurate and stable time-keeping. The best clocks
available today are atomic clocks, which lose less than a second in the age of the Universe. Better clocks lead to 
a host of other applications, including global positioning systems, space exploration, metrology and in fundamental physics.
Atomic clocks used to be confined in cells. But this results in a disturbance to the atoms when there are collisions with the cell walls.
This problem is circumvented by electromagnetic trapping, where the walls are replaced by 
a confining potential. An ideal clock would be an atom or ion at rest in space 
confined by static electromagnetic (and gravitational) fields. 
An obstacle to such confinement is  Earnshaw's theorem\cite{doi:10.1119/1.10449}: 
a static electric (or magnetic ) field satisfying the source-free Maxwell's equations cannot trap ions or atoms. 
Static electric fields are divergence and curl free and so satisfy the Laplace equation. The electrostatic potential is a harmonic
function and the maximum principle then tells us
that ions cannot be confined in a static electric field trap. Since static equilibrium is impossible, one uses dynamic electromagnetic fields
as in the Paul trap, where an ion is kept in place by an electric field configuration rotating at a radio frequency. 

Ions are charged therefore 
susceptible to disturbance from stray electromagnetic fields.
A neutral atom would be less susceptible to such fields but conversely, not as easy to 
manipulate.
If the atom has spin, it can be manipulated using
magnetic fields: a spinning atom in a magnetic field $\vec{B}$ has a potential energy $V_{dyn}(R)=-\mu | \vec{B}|$. 
However Wing's theorem tells us that this 
potential is also unable to confine atoms. It  can be shown \cite{wing,ketterle} that $-\nabla^2 V_{dyn}\ge 0$ and 
again the maximum principle forbids local minima of the potential.  Static confinement is impossible and so experimenters
use the magneto-optical trap in which electromagnetic fields at optical frequencies are used to confine the atom.

These general results are a consequence of the fact that laboratory electromagnetic fields  which are static have to satisfy Laplace's equation.
However geometric potentials can evade the theorems as they do not have to satisfy the source-free Maxwell's equations. 
The geometric electric field is curl-free (in regions where $|\vec{B}|$ is non-vanishing) but it need not be divergence free. We
have explicitly constructed examples of static trapping potentials using geometric fields. These examples clearly prove that
theorems analogous to Earnshaw's and Wing's theorem (and its generalisations\cite{ketterle})
do not apply when one includes geometric electromagnetic fields.

We describe two cases: static atom traps which confine neutral spinning 
atoms in three dimensions and atom wave-guides which confine them in two dimensions 
(say $x$ and $y$) and allow free motion in the $z$ direction. Neither traps nor wave-guides can be realised using static laboratory 
electromagnetic fields because of Wing's theorem. Both these can be achieved by using geometric electric fields. \\
{\it Geometric atom trap:} The laboratory magnetic field that is needed has two parts: a constant magnetic field $\vec{B}_0$ which is created by a solenoid
and an inhomogeneous magnetic field $\vec{B}_1(\vec{x})$ which is created by twelve wires located along the infinite line got by extending the 
edges of a cube of side $a$. A unit current is passed along these wires in the direction of the positive $x$, $y$ and $z$ directions. 
The total magnetic field is given by
\begin{equation}
\vec{B}(\vec{x})=\vec{B}_0+\vec{B}_1(\vec{x})
\end{equation}
The potential $V(\vec{x})$ seen by the atom is the sum 
\begin{equation}
V(\vec{x})=V_{dyn}(\vec{x})+V_{geom}(\vec{x})
\label{sum}
\end{equation}
of the dynamical potential  $V_{dyn}=-\mu |\vec{B}|$ and the geometric potential
$V_{geom}=1/4 \nabla_i \hat{ B}^a \nabla_i\hat{B}^a$.
We have used mathematica to compute the equipotentials of $V(\vec{x})$ and explicitly verified that it admits local minima.
The equipotentials in the neighbourhood of the local minima have spherical topology, 
rather than the hyperbolic surfaces one gets from saddle points. This proves that static magnetic fields can trap neutral atoms
if the geometric electric field is taken into account. 
\\
{\it Wave-guides:} To construct a geometric wave-guide we use $N$ straight 
wires along the $z$ direction equally spaced on the circumference of a circle of radius $a$ in the $x-y$ plane. These $N$ wires generate
the field $B_1(\vec{x})$, which is in the $x-y$ plane. In addition a solenoid generates a constant field $B_0$ in the $z$ direction. 
The resulting potential $V(\vec{x})$ (\ref{sum}) is plotted in Figure 1 for the case $N=10$. 
The equipotentials of $V(\vec{x})$ are topological circles  clearly showing two dimensional confinement. 
\begin{figure}
\begin{center}
\includegraphics[width=0.5\linewidth]{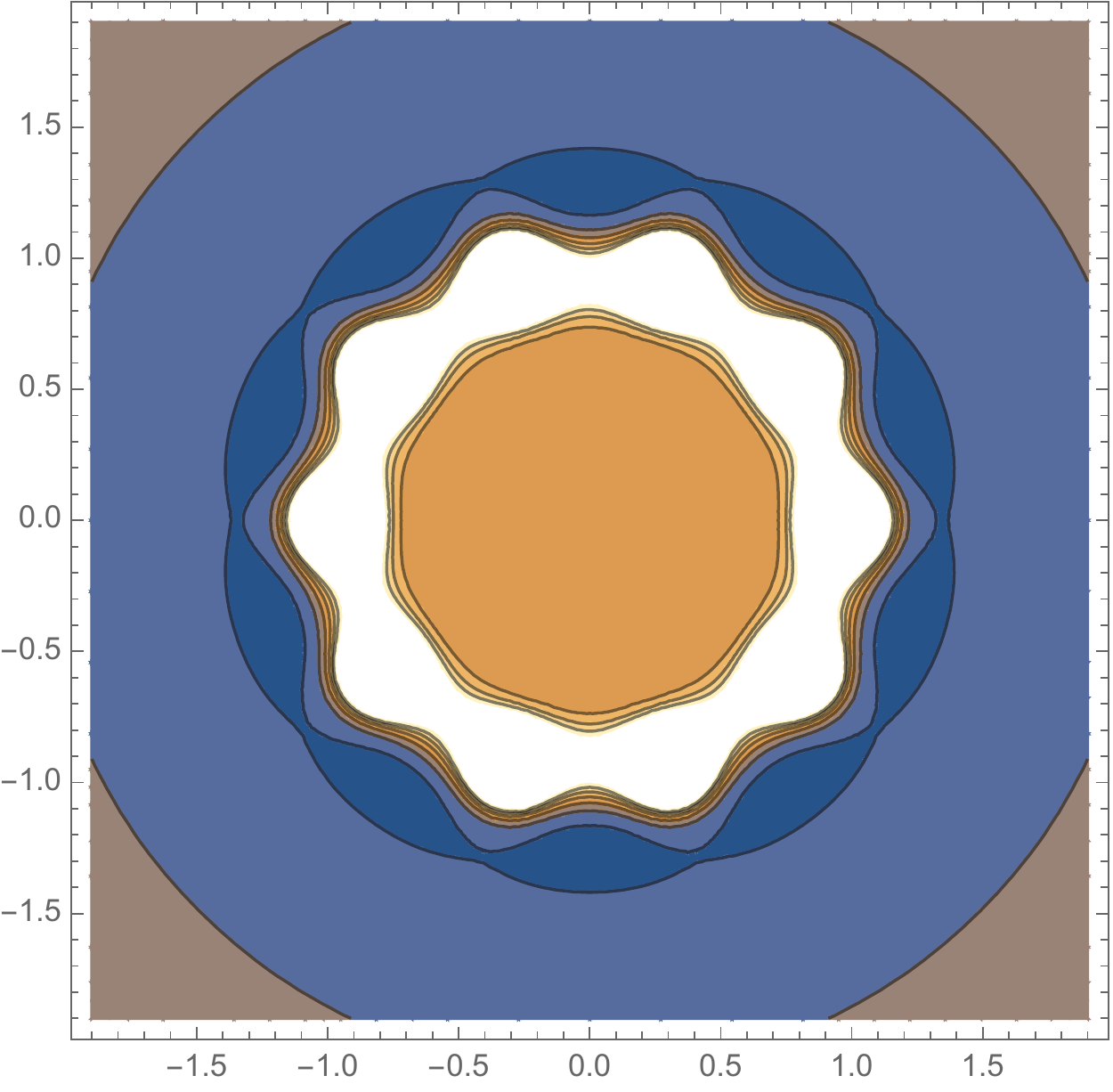}
\caption{Figure shows the equipotential lines (arbitrary units)  of the cross section of a geometric wave-guide. The inner most contours are of circular topology and contain a minimum of the potential $V(\vec{x})$. The potential rises from a minimum at the centre to a maximum in the white region and then falls
away in the radial direction. This shows that geometric fields can be used to evade Wing's theorem\cite{wing}.
\label{fig:lattice}}
\end{center}
\end{figure}
This proves that geometric fields can be used to evade the general no-go theorem of \cite{wing} and \cite{ketterle} 
preventing static trapping of neutral atoms.

Magneto-optical traps in current use are about a millimetre in size. Assuming that the Zeeman splitting due to
the magnetic field is $10^{10}{\textrm{Hz}}$, it is possible to roughly estimate $V_{dyn}$ as about $10^{-5}\rm{ev}$.  
$V_{geom}$ is about four orders smaller than this
and so entirely negligible in many current experiments. However, the situation changes dramatically if one reduces the size of the trap
to micron sizes. There is an ongoing effort to miniaturise atomic clocks \cite{Newman:19}, putting them on a chip. 
If one reduces the size of the trap by a factor
$f<1$, $V_{dyn}$ is unchanged (if the magnetic field remains the same), while $V_{geom}$ increases by a factor
of $1/f^2$. In smaller traps $V_{geom}$ is comparable in magnitude to $V_{dyn}$ and this would permit static trapping
of neutral atoms. Reducing the magnetic field decreases $V_{dyn}$, while leaving $V_{geom}$ unchanged. 
Thus, in going to smaller traps one {\it must} consider geometric electric fields and in fact, one can use them to advantage to 
achieve static trapping of neutral atoms\footnote{The adiabaticity condition is comfortably satisfied in the regime of interest.}.
The potential 
$V_{geom}$ describing the geometric electric field can trap neutral atoms without an optical field.

\section{Conclusion}
We have described a geometric effect that arises in the Born-Oppenheimer
approximation in a toy model . This is a scalar potential which has its origin in the 
geometry of ray space, namely the Fubini-Study metric.  We have given a magnetic field configuration in which
this geometric potential can be isolated from the other effects.
One difference between
$V_{geom}$ and the effects due to $V_{dyn}$ and $A_{geom}$ is that $V_{geom}$ has the same
effect on both spin states. But the other two act oppositely on the $\ket{+}$ and $\ket{-}$
states. The effect of $V_{geom}$ on neutrons is to favour regions where
the magnetic field direction is slowly varying and avoid regions of rapid 
variation\cite{MVBerry1990}, where the conditions of the adiabatic theorem would be violated.
 
While the more realistic models of molecular physics are not easy to solve,
the same geometric ideas apply there. If one solves the fast eigenvalue problem on
a computer, one can extract the slow dynamics in terms of eigenvalues and eigenstates
in an entirely geometric manner. The metric on the nuclear configuration space is
the one derived from the kinetic energy, which weights different nucleii according to their 
mass. The non-negativity of the geometric scalar potential remains true here
as it follows from general arguments. The magnitude of the geometric scalar potential
is smaller than the dynamical scalar potential by the ratio of the electronic mass to the nuclear mass. 
The effect of the geometric potential, while small, is 
measureable in the rotational and vibrational spectra of molecules.

In our toy model, we have treated the specific example of a neutron in a
strong magnetic field. However, the mathematics we presented here has
wider applications. One example is neutral cold atoms trapped using
magnetic fields\cite{pritchard}. Such configurations are used in cold atom laboratories to
achieve Bose-Einstein condensation as well as Fermi-degeneracy in dilute
atomic gases. Magnetic traps only confine atoms in the weak field-seeking
states. For weak magnetic field gradients and slow atoms, spin-flip transitions
are suppressed by the adiabatic theorem. In practice, these transitions do
occur in regions of very small magnetic fields which creates a region of trap
loss in the magnetic trap due to spin flips to untrapped states and are referred
to as Majorana losses. The background field ${\vec B}_0$ we used in our examples is designed to
avoid situations where the magnetic field is vanishingly small. We have also emphasised that the
geometric field will become more relevant as the size of traps becomes smaller, since it depends on gradients
of the direction of the magnetic field. This geometric field can also be used to advantage to develop better clocks.

Another application, unrelated to quantum mechanics, is polarised waves. When polarised light propagates through
a birefringent medium, the  medium acts like a magnetic field. In general one could have linear, circular and even
elliptical birefringence, in which two orthogonal polarisations traverse the medium at different speeds. Mathematically,
this problem  of polarised wave optics maps on all fours to the toy model of this paper. If one restricts to linear birefringence
and arranges for a helical configuration of molecules, one can even reproduce the example of section V in which the waves are affected only by the geometric
scalar potential. Such helical configurations can be generated as supramolecular structures using liquid crystals. The ideas described
here in the quantum context could be experimentally verified in the classical wave context of optics. The effect
of geometric potential can be seen in an interference experiment. This experiment could also be done in a scaled up setup by replacing light by radio waves and replacing the molecular helical variation with centimetre size
dielectric rods. One possible application is to make achromatic lenses using the geometric scalar potential to 
retard the beam. 

We hope to interest geometers as well as experimenters in understanding the uses of this geometric scalar potential.

{\it Acknowledgements}: This paper was motivated by a talk
given by Berry at ICTS, Bangalore, titled: The geometric phase and the separation of the world, during a program titled:
Geometric Phases in Optics and Topological Matter. I thank Sanjukta Roy for clearing up my 
ideas on atom traps and Supurna Sinha  for reading through the paper and 
suggesting improvements. This work was supported by a grant from the Simons Foundation (677895, R. G.).
\bibliographystyle{unsrt}
\bibliography{sample}
\end{document}